\documentclass{article}
\usepackage{microtype}
\usepackage{graphicx}
\usepackage{subcaption}
\usepackage{booktabs}
\usepackage{hyperref}

\usepackage[accepted]{icml2026}

\usepackage{amsmath}
\usepackage{amssymb}
\usepackage{mathtools}
\usepackage{amsthm}
\usepackage[capitalize,noabbrev]{cleveref}
\usepackage{enumitem}

\makeatletter
\renewcommand{\ICML@appearing}{\textit{Accepted at the ICML 2026 Workshop on
Machine Learning for Audio}. Copyright 2026 by the author(s).}
\makeatother

\icmltitlerunning{
    Probing Token Spaces under Generator Shift in AI-Generated Music Detection
}

\begin{document}

\twocolumn[
\icmltitle{Probing Token Spaces under Generator Shift in AI-Generated Music Detection}

\icmlsetsymbol{equal}{*}

\begin{icmlauthorlist}
  \icmlauthor{Joonyong Park}{utokyo,maap}
  \icmlauthor{Jungwoo Kim}{yonsei,maap}
  \icmlauthor{Junyoung Koh}{yonsei,maap}
  \icmlauthor{Yuki Saito}{utokyo}
\end{icmlauthorlist}

\icmlaffiliation{utokyo}{The University of Tokyo, Tokyo, Japan}
\icmlaffiliation{yonsei}{Yonsei University, Seoul, Republic of Korea}
\icmlaffiliation{maap}{MAAP Lab}

\icmlcorrespondingauthor{Joonyong Park}{joonyong-park@g.ecc.u-tokyo.ac.jp}

\icmlkeywords{AI-generated music detection, benchmark, neural audio codec, self-supervised audio, discrete tokens}

\vskip 0.3in
]

\printAffiliationsAndNotice{}

\begin{abstract}
AI-generated music detectors can appear robust on standard benchmark splits, yet their deployments require transfer to generator sources absent during training. We study this problem with source-restricted evaluation on \textsc{MoM-open}, an open reconstruction of MoM-CLAM that replaces the non-redistributable real corpus with FMA and MTG-Jamendo while preserving the fake-generator protocol. To isolate the role of representation, we introduce \textsc{CoMoE}, a compact fixed classifier for comparing heterogeneous audio token spaces while keeping the downstream architecture and training recipe unchanged. Experiments show that standard and real-source-restricted splits are nearly saturated, whereas fake-source restriction exposes large differences between token spaces: X-Codec tokens are strongest when training on Udio alone, while MERT-derived tokens are stronger when training on Suno-v3.5 alone. These results suggest that codec-style discrete token spaces should be treated as a primary experimental axis under generator shift in AI-generated music detection. Our code and data are available at \url{https://github.com/MAAP-LAB/CoMoE}.
\end{abstract}

\begin{figure*}[t]
  \centering
  \includegraphics[width=\textwidth]{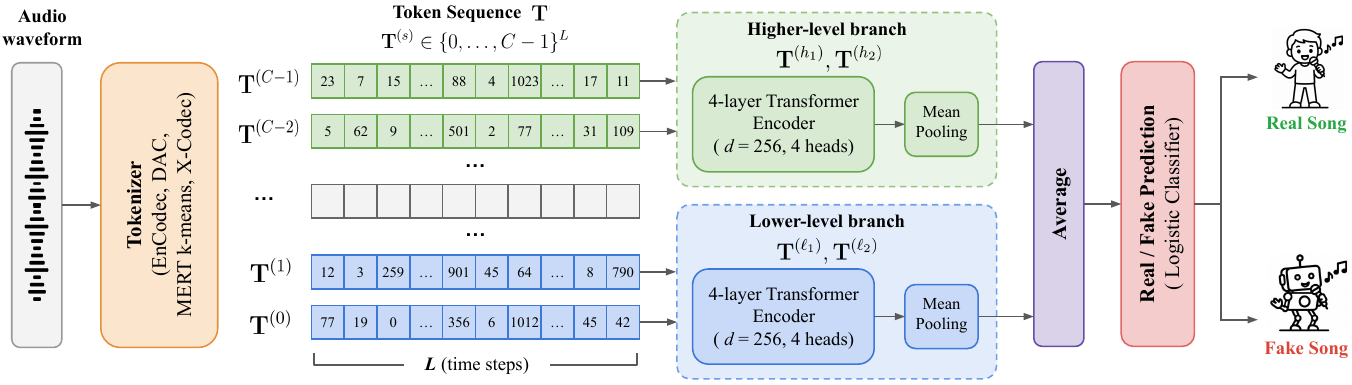}
  \caption{Architecture of \textsc{CoMoE}.}
  \label{fig:comoe_architecture}
  \vspace{-0.5mm}
\end{figure*}

\vspace {-2mm}\section{Introduction}\label{sec:intro}

AI-generated music detection aims to determine whether a music recording was produced by a human process or by a generative music system. The task is increasingly important as music generators can now produce full tracks with vocals, accompaniment, and near-release-quality production that are difficult to distinguish from human-made recordings~\citep{sonics,afchar2025aimusic,ai_music_arms_race,challenges_survey}. Recent detectors based on spectrograms, raw waveforms, and self-supervised audio representations report strong benchmark performance~\citep{mom_clam,sonics,afchar2025aimusic}. In deployment, however, a detector must flag outputs from generator sources that were absent during training, and standard benchmark splits may overstate this robustness when training and test sets share generator-specific artifacts~\citep{mom_clam,sonics,afchar2025aimusic}. This motivates not only source-restricted evaluation, but also a closer examination of which audio representations still transfer when generator-specific artifacts change.

  \vspace{-0.5mm}
In this work, we examine codec-style discrete audio tokens as candidates for transferable representations under generator shift. 
First, they provide a forensic view that differs from continuous acoustic or semantic features. Neural audio codecs represent audio as codebook sequences with residual-quantization structure~\citep{soundstream,encodec,dac}, which may expose codebook usage, token-transition, and quantizer-level patterns that are not directly isolated by pooled continuous features. Second, codec tokens provide a compact interface for downstream detectors: once tokens are extracted, the classifier can operate on symbolic sequences rather than full-resolution waveforms. Such codec structure has been explored in speech deepfake detection~\citep{safeear,codecfake,quantizer_aware_2026}, but music deepfake detection has mostly relied on waveform, spectrogram, or continuous representation detectors~\citep{sonics,mom_clam,afchar2025aimusic,fake_music_caps}. Importantly, codec tokens do not define a single representation: different tokenizers induce different discrete spaces, with different codebooks, temporal rates, and quantization behavior.

  \vspace{-0.5mm}
This variability makes tokenizer choice a key experimental variable rather than a preprocessing detail, especially under source-restricted evaluation.
To isolate this factor, we introduce Codec-Mixture-of-Experts (\textsc{CoMoE}), a compact fixed classifier for controlled tokenizer comparison. We keep the classifier architecture, training recipe, and evaluation protocol fixed, and replace only the input token space. We evaluate on \textsc{MoM-open}, an open reconstruction of MoM-CLAM that replaces the non-redistributable YouTube-derived real corpus with FMA-medium and MTG-Jamendo while preserving the fake-generator protocol. 

  \vspace{-0.5mm}
Our contributions are threefold: (i) we introduce \textsc{CoMoE} as a fixed classifier for comparing heterogeneous discrete audio token spaces;  (ii) construct \textsc{MoM-open} with source-restricted evaluation splits; and (iii) show that tokenizer choice is a primary experimental variable for cross-generator music deepfake detection.

\vspace {-2mm}\section{Related Work}\label{sec:related}

\textbf{Neural audio codecs and forensic cues.}
Neural audio codecs compress waveforms into compact latent or discrete token sequences for high-fidelity reconstruction~\citep{soundstream,encodec,dac}. Many modern codecs use residual vector quantization (RVQ), where audio is represented by multiple codebook streams that capture different levels of acoustic detail. In speech deepfake detection, neural-codec representations and quantizer hierarchies have already been used as forensic cues~\citep{safeear,codecfake,quantizer_aware_2026}. This suggests that codec tokens may reveal synthetic artifacts not directly exposed by continuous features.

\vspace{-0.5mm}
\textbf{Hybrid expert designs.}
Generated-content detectors often combine complementary views of the input. For example, AIDE uses both semantic and low-level artifact-sensitive branches for AI-generated image detection~\citep{clip,aide}. This motivates branch-specialized designs for codec-token detection, where different codebook levels may carry different forensic information. For music deepfake detection, however, the unresolved question is not only how to design a classifier, but whether the token space itself controls robustness to unseen generators. 

\vspace{-0.5mm}
\textbf{Music deepfake detection.}
Music deepfake detectors mostly rely on raw waveforms, spectrograms, or continuous self-supervised features. SONICS uses temporal and spectral tokenization over mel-spectrograms~\citep{sonics}, while CLAM uses continuous MERT and Wav2Vec2 streams~\citep{mom_clam}. Other studies similarly evaluate raw-audio, spectrogram, or pretrained-representation baselines~\citep{afchar2024detecting,fake_music_caps}. In contrast, codec-style discrete token spaces have not been systematically compared under cross-generator music deepfake evaluation.

\vspace {-2mm}\section{\textsc{CoMoE}: A Controlled Token-Space Probe}\label{sec:model}

\vspace{-0.5mm}
\textbf{Architecture.}
\Cref{fig:comoe_architecture} explains the structure of the model overall. The four streams consist of two lower-level and two higher-level token streams. This is a controlled interface rather than a theoretical constraint: for RVQ codecs, the streams are selected from early and late codebooks; for MERT $k$-means, they are selected from lower and upper self-supervised layers.

Formally, \textsc{CoMoE} consumes four discrete token streams,
\begin{equation}
\begin{gathered}
\mathbf{T}
=
\left(
\mathbf{T}^{(\ell_1)},
\mathbf{T}^{(\ell_2)},
\mathbf{T}^{(h_1)},
\mathbf{T}^{(h_2)}
\right),\\
\mathbf{T}^{(s)} \in \{0,\dots,C-1\}^{L},
\end{gathered}
\end{equation}
where $C$ is the codebook size, $L$ is the fixed token sequence length after truncation or padding, and $s$ indexes one of the four streams. The superscripts $\ell$ and $h$ denote lower- and higher-level streams, respectively. The two lower-level streams $\mathbf{T}^{(\ell_1)},\mathbf{T}^{(\ell_2)}$ and the two higher-level streams $\mathbf{T}^{(h_1)},\mathbf{T}^{(h_2)}$ are processed by separate Transformer encoders $f^{(\ell)}$ and $f^{(h)}$ with identical architecture. Each encoder has four layers, hidden size $d=256$, and four attention heads.

The encoder outputs are mean-pooled over time to obtain two branch representations,
\begin{equation}
\begin{aligned}
\mathbf{h}^{(\ell)}
&=
\mathrm{Pool}\!\left(
f^{(\ell)}
\left(
\mathbf{T}^{(\ell_1)},
\mathbf{T}^{(\ell_2)}
\right)
\right),\\
\mathbf{h}^{(h)}
&=
\mathrm{Pool}\!\left(
f^{(h)}
\left(
\mathbf{T}^{(h_1)},
\mathbf{T}^{(h_2)}
\right)
\right),
\end{aligned}
\end{equation}
where $\mathbf{h}^{(\ell)},\mathbf{h}^{(h)} \in \mathbb{R}^{d}$. The two branch representations are averaged and fed to a binary logistic classifier:
\begin{equation}
\mathbf{z}
=
\frac{1}{2}
\left(
\mathbf{h}^{(\ell)}
+
\mathbf{h}^{(h)}
\right),
\quad
\hat{y}
=
\sigma\!\left(\mathbf{w}^{\top}\mathbf{z}+b\right),
\end{equation}
where $\mathbf{w}\in\mathbb{R}^{d}$ and $b\in\mathbb{R}$ are trainable classifier parameters. All \textsc{CoMoE} variants use the same four-stream classifier, so differences among \textsc{CoMoE} rows reflect the input token space rather than changes in the downstream classifier.

\vspace{-2mm}
\textbf{Token front-ends.}
All tokenizers are mapped to the fixed four-stream interface defined above, with codebook size $C=1024$ and fixed sequence length $L$ after truncation or padding. For each tokenizer, we construct two lower-level streams $\mathbf{T}^{(\ell_1)},\mathbf{T}^{(\ell_2)}$ and two higher-level streams $\mathbf{T}^{(h_1)},\mathbf{T}^{(h_2)}$.

\vspace{-1.5mm}
\textbf{EnCodec 24 kHz}~\citep{encodec} provides acoustic RVQ codebook streams\footnote{\url{huggingface.co/facebook/encodec_24khz}}. We map codebooks $q=0,1$ to $\mathbf{T}^{(\ell_1)},\mathbf{T}^{(\ell_2)}$ and codebooks $q=6,7$ to $\mathbf{T}^{(h_1)},\mathbf{T}^{(h_2)}$.

\vspace{-1.5mm}
\textbf{DAC 44 kHz}~\citep{dac} is used as a second acoustic codec\footnote{\url{github.com/descriptinc/descript-audio-codec}}. We apply the same early/late rule and map codebooks $q=0,1$ to $\mathbf{T}^{(\ell_1)},\mathbf{T}^{(\ell_2)}$ and codebooks $q=7,8$ to $\mathbf{T}^{(h_1)},\mathbf{T}^{(h_2)}$.

\vspace{-1.5mm}
\textbf{X-Codec mini}~\citep{xcodec} is a music-trained semantic-aware codec checkpoint\footnote{\url{huggingface.co/m-a-p/xcodec_mini_infer}}. X-Codec mini provides twelve RVQ codebook streams; we map codebooks $q=0,1$ to $\mathbf{T}^{(\ell_1)},\mathbf{T}^{(\ell_2)}$ and codebooks $q=10,11$ to $\mathbf{T}^{(h_1)},\mathbf{T}^{(h_2)}$.

\vspace{-1mm}
To compare neural audio codec tokens with self-supervised discrete units, we also construct \textbf{MERT $k$-means} tokens from MERT-v0-public hidden states~\citep{mert}\footnote{\url{huggingface.co/m-a-p/MERT-v0-public}}. We use layers $\{0,1,11,12\}$, cluster frame features with MiniBatch $k$-means~\citep{sculley_minibatch}, and emit four streams of discrete units. Layers $0,1$ are mapped to $\mathbf{T}^{(\ell_1)},\mathbf{T}^{(\ell_2)}$, and layers $11,12$ are mapped to $\mathbf{T}^{(h_1)},\mathbf{T}^{(h_2)}$.

\begin{table}[t]
\caption{\textsc{MoM-open} composition. Real audio is drawn from two openly redistributable music corpora; fake audio follows the $\mathcal{F}_T$/$\mathcal{F}_O$ designation of the original benchmark~\citep{mom_clam}.}
\label{tab:dataset}
\vskip 0.05in
\centering
\footnotesize
\setlength{\tabcolsep}{4pt}
\begin{tabular}{llr}
\toprule
Class & Source & Clips \\
\midrule
Real $\mathcal{R}$ & FMA-medium~\citep{fma} & $24{,}979$ \\
(train+test) & MTG-Jamendo~\citep{mtgjamendo} & $52{,}501$ \\
\cmidrule(lr){1-3}
Fake $\mathcal{F}_{\rm T}$ & Suno-v2~\citep{suno} & $660$ \\
(train) & Suno-v3.5~\citep{suno} & $28{,}611$ \\
& Udio~\citep{udio} & $19{,}500$ \\
& DiffRhythm~\citep{diffrhythm} & $4{,}594$ \\
\cmidrule(lr){1-3}
Fake $\mathcal{F}_{\rm O}$ & Riffusion~\citep{riffusion} & $7{,}043$ \\
(OOD) & Suno-v3~\citep{suno} & $3{,}116$ \\
& Suno-v4~\citep{suno} & $27$ \\
& YuE~\citep{yue} & $5{,}278$ \\
\midrule
\textbf{Total} & & \textbf{$146{,}309$} \\
\bottomrule
\end{tabular}
\vskip -0.1in
\end{table}

\vspace {-0.5mm}\textbf{Continuous MERT ablation.}
To separate the effect of MERT representations from the effect of discretization, we also evaluate a continuous-input ablation. This model uses the same low/high Transformer backbone as \textsc{CoMoE}, but replaces the token embedding lookup with a linear projection of continuous MERT-v0 frame features. We use the same four MERT layers, $\{0,1,11,12\}$, mapping layers $0,1$ to the lower-level branch and layers $11,12$ to the higher-level branch. This variant is not a discrete-token \textsc{CoMoE} model; it is included only to test whether the MERT $k$-means result is due to discretization or to the underlying MERT representation.

\vspace {-0.5mm}\textbf{Baselines.}
We include two non-\textsc{CoMoE} baselines. \textsc{MLP} (MERT) uses mean-pooled MERT-v0-public features followed by a small multilayer perceptron. \textsc{CLAM}~\citep{mom_clam} is the dual-rate reference detector from the original benchmark, using MERT and Wav2Vec2 streams with weighted cross-attention. The MERT-MLP and CLAM baselines follow their respective recipes.

\vspace {-0.5mm}\textbf{Training.}
All \textsc{CoMoE} variants are trained with the same recipe: 12 epochs of AdamW~\citep{AdamW}, learning rate $2\times10^{-4}$, label smoothing 0.05, seed 42, and a single H100 GPU. The MERT-MLP and CLAM baselines follow their respective baseline recipes.

\begin{table}[t]
\caption{Split definitions. The held-out target additionally contains the base-split OOD real set so AUC is computed in the standard binary sense.}
\label{tab:splits}
\vskip 0.05in
\centering
\footnotesize
\setlength{\tabcolsep}{4pt}
\begin{tabular}{lll}
\toprule
Split & Train & Test \\
\midrule
base & $\mathcal{R}$ train $\cup$ $\mathcal{F}_{\rm T}$ & $\mathcal{R}$ test $\cup$ $\mathcal{F}_{\rm O}$ \\
Real-FMA & \textbf{FMA} $\cup$ $\mathcal{F}_{\rm T}$ & $\mathcal{F}_{\rm O}$ $\cup$ Jamendo \\
Real-Jamendo & \textbf{Jamendo} $\cup$ $\mathcal{F}_{\rm T}$ & $\mathcal{F}_{\rm O}$ $\cup$ FMA \\
Fake-Suno3.5 & $\mathcal{R}$ train $\cup$ \textbf{Suno-v3.5} & $\mathcal{R}$ test $\cup$ ($\mathcal{F}$ $\setminus$ Suno-v3.5) \\
Fake-Udio & $\mathcal{R}$ train $\cup$ \textbf{Udio} & $\mathcal{R}$ test $\cup$ ($\mathcal{F}$ $\setminus$ Udio) \\
\bottomrule
\end{tabular}
\vspace {-4mm}
\vskip -0.1in
\end{table}

\begin{table*}[t]
\caption{OOD AUC (\%) on \textsc{MoM-open} across the base split, real-source-restricted splits, and fake-source-restricted splits. Split names indicate the source retained in training. Values in parentheses are absolute changes from the corresponding base AUC in percentage points.}
\label{tab:main}
\vskip 0.05in
\centering
\small
\setlength{\tabcolsep}{4pt}
\begin{tabular}{lccccc}
\toprule
Model & base & \textsc{Real-FMA} & \textsc{Real-Jamendo} & \textsc{Fake-Suno3.5} & \textsc{Fake-Udio} \\
\midrule
\textsc{CLAM} & 99.92 & \textbf{99.71} ($-0.2$) & \textbf{99.85} ($-0.1$) & \textbf{97.72} ($-2.2$) & 66.51 ($-33.4$) \\
\textsc{MLP} (MERT) & 99.77 & 99.07 ($-0.7$) & 99.47 ($-0.3$) & 86.87 ($-12.9$) & 67.45 ($-32.3$) \\
\midrule
\textsc{CoMoE} (X-Codec) & \textbf{99.93} & 99.62 ($-0.3$) & 99.73 ($-0.2$) & 86.97 ($-13.0$) & \textbf{89.04} ($-10.9$) \\
\textsc{CoMoE} (DAC) & 99.82 & 98.98 ($-0.8$) & 99.51 ($-0.3$) & 88.33 ($-11.5$) & 77.28 ($-22.6$) \\
\textsc{CoMoE} (EnCodec) & 96.44 & 95.64 ($-0.8$) & 94.76 ($-1.7$) & 85.15 ($-11.3$) & 58.64 ($-37.8$) \\
\textsc{CoMoE} (MERT $k$-means) & 99.83 & 99.14 ($-0.7$) & 99.53 ($-0.3$) & 92.22 ($-7.6$) & 73.26 ($-26.6$) \\
\midrule
\textsc{MERT-continuous} (same backbone) & 99.87 & 99.01 ($-0.9$) & 99.57 ($-0.3$) & 93.84 ($-6.0$) & 71.91 ($-28.0$) \\
\bottomrule
\end{tabular}
\vskip -0.1in
\end{table*}

\begin{table}[t]
\caption{Held-out-fake detection rate (\%) under the validation-selected threshold.}
\label{tab:fake_detection_rate}
\vskip -0.05in
\centering
\small
\setlength{\tabcolsep}{5pt}
\begin{tabular}{lcc}
\toprule
Model & \textsc{Fake-Suno3.5} & \textsc{Fake-Udio} \\
\midrule
\textsc{CLAM}                       & \textbf{71.0} & 2.6 \\
\textsc{MLP} (MERT)                 & 60.1 & 26.0 \\
\textsc{CoMoE} (X-Codec)            & 38.7 & \textbf{45.1} \\
\textsc{CoMoE} (EnCodec)            & 43.8 & 23.5 \\
\textsc{CoMoE} (DAC)                & 61.4 & 29.2 \\
\textsc{CoMoE} (MERT $k$-means)     & 51.9 & 17.3 \\
\textsc{MERT-continuous} & 49.9 & 7.8 \\
\bottomrule
\end{tabular}
\vskip -0.1in
\end{table}

\vspace {-2mm}
\section{\textsc{MoM-open} and Source-Restricted Splits}
\label{sec:momopen}

\textbf{Dataset.}
We construct \textsc{MoM-open}, an open reconstruction of MoM-CLAM. Since the original benchmark relies on YouTube-derived real audio that is difficult to redistribute or reliably rebuild, we replace the real half with FMA-medium and MTG-Jamendo while keeping the original fake-generator protocol. These corpora have been widely used for music information retrieval and audio-based music analysis tasks, including tagging, genre analysis, and popularity prediction~\citep{fma,mtgjamendo,lee2018music}. \Cref{tab:dataset} summarizes the resulting 146{,}309 clips. All clips are normalized to a shared audio representation by standardizing duration, channel configuration, sampling rate, codec, and metadata handling.

\vspace{-0.5mm}
\textbf{Source-restricted splits.}
\Cref{tab:splits} defines the evaluation splits. The base split follows the original fake-generator partition. Real-source restriction tests whether detectors rely on FMA- or Jamendo-specific real-corpus cues, while fake-source restriction tests whether a detector trained on one fake generator source transfers to unseen fake sources.

\vspace{-3mm}
\section{Results}\label{sec:results}

\vspace{-1mm}
\noindent\textbf{Metrics and validation.}
For each condition, validation examples are drawn only from the sources retained in the training split; held-out fake sources are never used for threshold selection. We report AUC and held-out-fake detection rate. The latter uses a threshold $\tau^{\star}$ selected by maximizing validation F1 and then applied unchanged to each held-out generator source.

\vspace{-1mm}
\noindent\textbf{Base and real-source-restricted splits are nearly saturated.}
\Cref{tab:main} shows that the base split is close to saturated for all strong detectors: \textsc{CLAM}, \textsc{MLP} (MERT), and most \textsc{CoMoE} variants reach AUCs near 99.8--99.9\%, except for the lower but still high EnCodec-token \textsc{CoMoE}. Real-source restriction is also mild, with much smaller drops than the fake-source-restricted conditions.

\vspace{-1mm}
\noindent\textbf{Fake-source restriction exposes large model differences.}
The rightmost two columns of \Cref{tab:main} are much more discriminative than the base or real-source-restricted splits. In \textsc{Fake-Suno3.5}, \textsc{CLAM} remains strongest. In \textsc{Fake-Udio}, however, \textsc{CLAM} drops sharply, while \textsc{CoMoE} with X-Codec tokens becomes the strongest configuration.

\vspace{-1mm}
\noindent\textbf{Token-space identity is the dominant factor among fixed-architecture \textsc{CoMoE} variants.}
Because all \textsc{CoMoE} rows in \Cref{tab:main} use the same classifier, their differences isolate the input token space. Under \textsc{Fake-Udio}, EnCodec drops to 58.64\%, DAC improves over EnCodec but remains below X-Codec, and X-Codec reaches 89.04\%. MERT $k$-means is strongest among \textsc{CoMoE} variants on \textsc{Fake-Suno3.5}, whereas X-Codec is strongest on \textsc{Fake-Udio}.

\vspace{-1mm}
\noindent\textbf{Pooled MERT features alone are not sufficient.}
The \textsc{MLP} (MERT) baseline in \Cref{tab:main} tests whether mean-pooled continuous music-SSL features alone explain the robustness gains. Although it is strong on the base and real-source-restricted splits, it drops substantially under fake-source restriction, especially \textsc{Fake-Udio}. Thus, the X-Codec result cannot be explained simply by using a music-pretrained representation; sequential token structure also matters.

\vspace{-1mm}
\noindent\textbf{Discretization alone does not explain AUC, but affects operating-point stability.}
The \textsc{MERT-continuous} ablation in \Cref{tab:main} uses the same low/high Transformer backbone as MERT $k$-means, but replaces discrete units with continuous MERT frame features. It improves AUC on \textsc{Fake-Suno3.5}, but is slightly worse on \textsc{Fake-Udio}; thus, AUC differences are not explained by discretization alone. However, \Cref{tab:fake_detection_rate} shows a larger operating-point gap: under \textsc{Fake-Udio}, MERT $k$-means retains 17.3\% held-out-fake detection rate, while MERT-continuous drops to 7.8\%.

\vspace{-1mm}
\noindent\textbf{AUC and operating-point behavior diverge.}
\Cref{tab:fake_detection_rate} shows that validation-selected thresholds do not always transfer to held-out fake sources. The clearest case is \textsc{CLAM}: under \textsc{Fake-Udio}, it retains non-random AUC in \Cref{tab:main}, but its held-out-fake detection rate drops to 2.6\%. In contrast, \textsc{CoMoE} with X-Codec tokens gives the best \textsc{Fake-Udio} detection rate and the smallest cross-condition gap, suggesting that fake-source restriction should be evaluated with both ranking and operating-point metrics.

\vspace{-4mm}
\section{Conclusion}\label{sec:conclusion}
\vspace{-2mm}

We presented a controlled study of cross-generator AI-generated music detection in which the downstream classifier is fixed and only the audio token space is varied. Experiments on \textsc{MoM-open} show that standard and real-source-restricted splits are nearly saturated, while fake-source restriction reveals large differences between token spaces. These results suggest that codec-style discrete token spaces should be treated as a primary experimental axis in music deepfake detection, rather than as a preprocessing detail. However, the study has some limitations: \textsc{MoM-open} is an open reconstruction, and X-Codec mini is not lineage-free with respect to YuE-related tooling. Future work should evaluate more generator sources, control training-pool size, and test calibration or fusion methods under generator shift.

\textbf{Acknowledgement.} This work was supported by JSPS KAKENHI Grant Number 26KJ0771. 

\bibliography{example_paper}
\bibliographystyle{icml2026}

\end{document}